\documentclass[superscriptaddress,twocolumn,prl]{revtex4-2}

\usepackage{graphicx}%
\usepackage{amsmath,amssymb,amsfonts}%
\usepackage{amsthm}%
\usepackage{mathrsfs}%
\usepackage{xcolor}%
\usepackage{braket}%

\theoremstyle{thmstyleone}%

\theoremstyle{thmstyletwo}%

\theoremstyle{thmstylethree}%

\newcommand{\red}[1]{{\color{black}#1}}

\begin{document}
\title{Experimental visual quantum state discrimination}%

\author{A. Duplinskiy}\email{al.duplinskiy@gmail.com}
\affiliation{Department of Physics, University of Oxford, Oxford, OX1 3PU, UK}
\author{M. J. Filipovich}
\affiliation{Department of Physics, University of Oxford, Oxford, OX1 3PU, UK}
\author{A. I. Lvovsky}
\affiliation{Department of Physics, University of Oxford, Oxford, OX1 3PU, UK}

\begin{abstract}
Almost all solutions to vision tasks --- offered by nature and technology alike --- rely on capturing an image using a lens or a similar optical instrument, followed by computational analysis by neurons of a brain or a digital computer. This approach is inherently suboptimal in photon-limited and noisy conditions, where critical information is irreversibly lost at the measurement stage. We address this gap by introducing AI-driven optimization of quantum measurements on the incoming photons, learning how
to extract the maximum possible information for a given computer vision task from each incoming photon. The trained measurement consists of %
a diffractive optical neural network (DONN) followed by a detector array. %
We train the DONN with a quantum information-theoretic objective that maximizes the rate at which the discrimination error decays as photons accumulate. The resulting measurement achieves exponentially lower error than both direct imaging and a DONN optimized for single-photon discrimination. We validate the advantage in a noiseless setting as well as a photon-counting experiment with an electron-multiplying CCD with the noise corresponding to $10^4$ photons per single signal photon, where the measurement also learns to mitigate the detector noise.
\end{abstract}

\maketitle

\section*{Introduction}

Computer vision now matches or exceeds human performance on a wide range of
recognition and inference tasks \cite{lecun2015deep,he2015delving}. Almost
without exception, these systems use direct imaging (DI), in which the scene is mapped onto a camera sensor to estimate its spatial intensity distribution. This approach is justified for high-intensity illumination, in which so many photons reach the sensor that the optical field can be treated as a classical wave. However, light is not abundant in many visual AI scenarios. Examples include live cell fluorescence microscopy, in which the intensity is capped by photobleaching and phototoxicity \cite{icha2017phototoxicity} or dark-sky astronomy and long-range single-photon LIDAR, where large propagation distances severely decrease the amount of available light \cite{kirmani2014first}. Under photon-starved conditions, the image data become contaminated by the shot noise arising from discrete, quantum nature of light, as well as the detection noise of the camera. The performance of conventional vision pipelines degrades sharply as the photon budget shrinks.

The conventional response to this challenge has been to develop better post-processing algorithms, including denoising, photon-counting reconstruction, and learned image priors \cite{shin2015photon}. Yet, these methods operate only after the state of the optical field has already been measured, %
resulting in some of the information carried by that state getting irreversibly lost in the measurement process. 
We address this gap by moving intelligence further upstream: into the measurement itself. %
This can be helpful because for many computer-vision tasks, including classification, object detection and regression, reconstructing the image is not the ultimate goal. A measurement can instead be optimized specifically for the inference task at hand.

We focus on classification of visual scenes. Each $i$th class consists of $M$  instances, each of which emits photons in a coherent spatial state $\ket{\psi_{ij}}$. The task is, given $N$ photons each coming from an arbitrary instance of a specific class, to determine which class it is. This is equivalent to discrimination of quantum states  \cite{barnett2009quantum} represented by density operators 
\begin{equation}\label{rho}
    \hat\rho_i=\frac{1}{M}\sum_j\ket{\psi_{ij}}\bra{\psi_{ij}},
\end{equation}
based on $N$ separable (single-copy) measurements. Framing the task this way makes the measurement the sole object of optimization and supplies fundamental benchmarks. Helstrom's detection theory identified the minimum-error measurement for two known states \cite{helstrom1976quantum}. When multiple copies of each state are available, the  Helstrom bound becomes difficult to calculate, but is asymptotically approximated \red{in the limit $N\to\infty$} by the quantum Chernoff bound \cite{audenaert2007discriminating}. This bound can be reached by collective (entangling) measurements, which are difficult to implement experimentally. For single-copy measurements, such as in our case, \red{a tighter asymptotic estimate for $N\to\infty$ is provided by the Fuchs-Caves bound \cite{fuchs1995mathematical}. It is also known that for finite $N$,} the optimal measurement depends on the number of copies available \cite{acin2005multiple}. %

Research on quantum state discrimination discovered various strategies for constructing measurements that can be optimal in different scenarios \cite{calsamiglia2010local, higgins2011multiple, dolinar1973optimum, cook2007optical, becerra2013experimental, jagannathan2022demonstration}. Most of this research, however, concerned with relatively simple low-dimensional systems, which allow analytic or numerical optimization. But this approach becomes prohibitive for real-world multi-dimensional scenarios associated with computer vision. Our innovation, explored theoretically in our recent work \cite{filipovich2026learned} and experimentally here, consists in designing the measurement via machine learning methods. We use these methods to optimize (``train") the measurement to extract the maximum possible information from each incoming photon. %

The primary tool in our method is the diffractive optical neural network (DONN) --- a sequence of learned phase masks that passively reshape the optical field before it reaches a detector array. The DONN applies a unitary transformation to the incoming light field, which, along with position-sensitive detection, implements a trainable optical measurement.  Since their inception in 2018  \cite{lin2018alloptical}, diffractive ONNs have been used for a variety of applications including communications, image recognition and reconstruction \cite{chen2024}. Although a DONN can realize only a constrained subset of possible measurements, it spans a large, trainable parameter space within which the measurement basis can be optimized toward the best available one for the task.

Our approach has been inspired by a decade of research on quantum-optimal sensing under diffraction-limited conditions \cite{tsang2016quantum,lvovsky2026passive}. To estimate simple parameters of a scene --- such as the separation of a sub-Rayleigh point source pair --- it suffices to project the field onto a fixed, generic mode basis. DONNs have successfully been used for this task \cite{buonaiuto2025machine, li2024first,rouviere2024ultra}, with the training aimed at optimizing the fidelity of mode sorting into that basis. However, for computer vision tasks involving complex scenes, the intermediate step of finding the best measurement basis can be omitted and the DONN can be trained directly to maximize the final accuracy \cite{filipovich2026learned}.

Optical neural networks have been used for visual classification, in settings involving active interaction with the sample. Ma et al.~\cite{ma2026machine} used a phase plate to optimize the illumination of the scene and classify by detecting the reflected light. Classification using Hong–Ou–Mandel interference between a single photon representing the image and a probe photon shaped by a trained diffractive element was reported by Minati {\it et al.} \cite{minati2026}. Complementing this, Goel {\it et al.} demonstrated an experimental scheme for quantum state discrimination using a four-layer multiplane light converter capable of resolving superpositions of Hermite–Gaussian modes~\cite{goel2023}. Basic machine learning by means of a quantum ONN has been demonstrated by Yin {\it et al.}, albeit outside the sensing context \cite{yin2025}. A general paradigm of quantum-computational sensing and the promises of this approach have been laid out by Khan {\it et al.} \cite{khan2025}.

\section*{Optical neural network and its training}

We compare DONNs trained with two different objectives. The first type --- \textit{single-photon-optimized} measurement -- is aimed at maximizing accuracy when a measurement on just one photon is performed~\cite{fan2026quantum}. The second --- \textit{scaling-optimized} measurement -- targets the \emph{rate} at which the accuracy improves with the growing number of detected photons. %
To define these objectives quantitatively, we formalize the DONN as a unitary transformation $\hat V(\Phi)$, where $\Phi$ is the parameter set defining the phase masks, on the input quantum state $\hat\rho_i$. The DONN output is measured by an array of photon detectors. The probability of an event in a detector localized at $x$ is then $P_i(x|\Phi)=\bra{x} \hat V\hat\rho_i\hat V^\dagger\ket x$. 

When only one photon is available, and detected at $x$, the classification inference is made according to the class most likely to send a photon there. The probability to correctly classify a photon of the $i$th class is therefore 
\begin{equation}
A_i(\Phi) =  \sum_{x:\forall j\ne i\,P_j(x|\Phi)<P_i(x|\Phi)} \, P_i(x \mid \Phi).
\label{eq:single_photon_acc_0}
\end{equation}
For $K$ equiprobable classes the probability of correct classification simplifies to
\begin{equation}
A(\Phi) = \frac{1}{K} \sum_i A_i(\Phi)= \frac{1}{K} \sum_{x} \max_i \, P_i(x \mid \Phi).
\label{eq:single_photon_acc}
\end{equation}
This probability is the training objective for the single-photon-optimized case. In practice, we train by minimizing $-\log A(\Phi)$, replacing the hard maximum with a per-pixel softmax over classes to keep the objective differentiable.

Single-photon-optimized measurements do not perform best for multiple photons, even if the measurements on individual photons are separable \cite{acin2005multiple, higgins2009mixed, higgins2011multiple}. An alternative strategy is therefore to train the DONN to construct the measurement that optimizes the exponential classification error decay rate in the asymptotic limit of infinitely many photons, which for $K=2$ becomes the classical Chernoff information, 
\begin{equation}
\xi(\Phi) = -\min_{s \in [0,1]} \log \left[ \sum_{x} P_1(x \,|\, \Phi)^s \, P_2(x \,|\, \Phi)^{1-s} \right].
\label{eq:chernoff}
\end{equation}
In practice, the optimal $s$ in Eq.~\eqref{eq:chernoff} lies at $s \simeq 1/2$. %
We found that fixing $s=1/2$ produced no measurable loss in performance while speeding up training, and therefore adopt this setting throughout the study. For this setting, the objective function  \eqref{eq:chernoff} becomes the Bhattacharyya distance 
\begin{equation}
B(\Phi) = -\log \left[ \sum_{x} \sqrt{P_1(x|\Phi)\, P_2(x|\Phi)} \right]
\label{eq:bhattacharyya}
\end{equation}

\red{The maximum of Eq.~(\ref{eq:bhattacharyya}) for single-copy measurements is given by the Fuchs-Caves bound: $B(\Phi) \le B_\mathrm{FC} = -\log \mathrm{tr}\sqrt{\hat\rho_2^{1/2}\hat\rho_1\hat\rho_2^{1/2}}$, attained by a projective measurement in the eigenbasis of $\hat M = \hat\rho_2^{-1/2}\sqrt{\hat\rho_2^{1/2}\hat\rho_1\hat\rho_2^{1/2}}\hat\rho_2^{-1/2}$ \cite{fuchs1995mathematical}. This bound corresponds to the Bhattacharyya setting ($s=1/2$) and therefore not strict. However, since we found this setting to be very close to optimal, the Fuchs-Caves bound provides a good approximation for the maximum attainable performance when collective measurements cannot be carried out.}

We show in Methods that the accuracy of DONNs trained with this objective, with sufficient depth of the DONN and properly accounting for optical losses and detector noise, %
approaches the fundamental limit set by the quantum Chernoff information. 

The trained DONN is used for inference. The inference is attained following the straightforward maximum-likelihood rule: for any observed sample, the best guess is the state under which that sample is most likely. The likelihood is given by
\begin{equation}
\label{eq:likelihood}
\mathcal{L}_i(\mathbf{n})
=
\prod_{x} P_i(x \,|\, \Phi)^{\,n_x},
\end{equation}
where $\mathbf n=\{n_x\}$ is the vector of photon numbers detected in all pixels.

\section*{Results}

\begin{figure}
\centering
\includegraphics[width=0.95\linewidth]{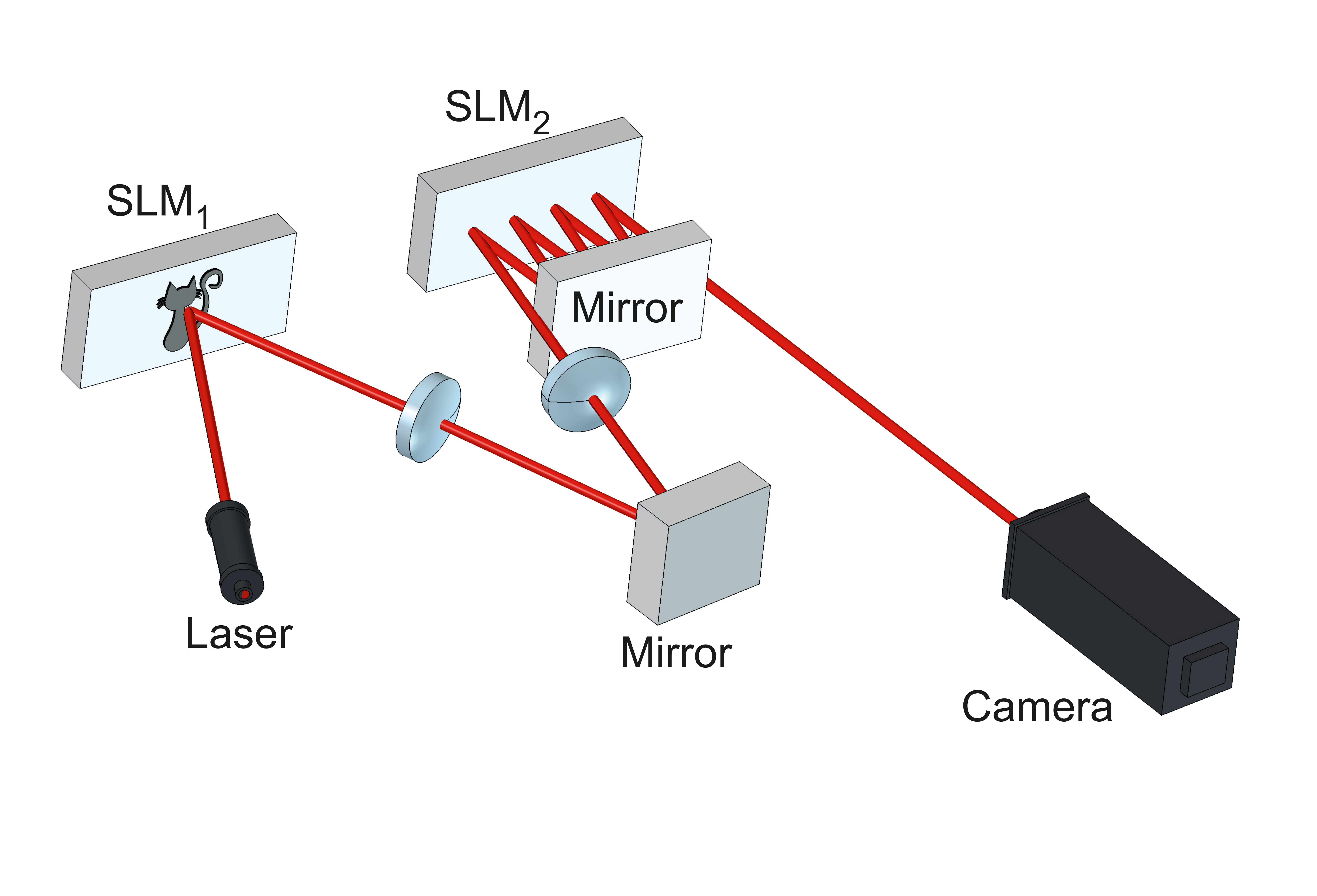}
\caption{Experimental setup features two SLMs, the first one creating an object and the second implementing the DONN with four phase plates. Output light is detected via a classical camera or an EM-CCD.}
\label{fig:setup_simple}
\end{figure}

In the experiment (Fig.~\ref{fig:setup_simple}), each object is encoded into the amplitude of a coherent optical field by a spatial light modulator (SLM). The field then propagates through the trainable measurement -- a four-layer DONN realized by a second SLM together with a mirror, displaying the phase masks -- and is finally detected by a camera. The masks are trained offline on a digital twin of the setup; further details are given in Methods.

\subsection*{Binary classification}

\begin{figure}
\centering
\includegraphics[width=1\linewidth]{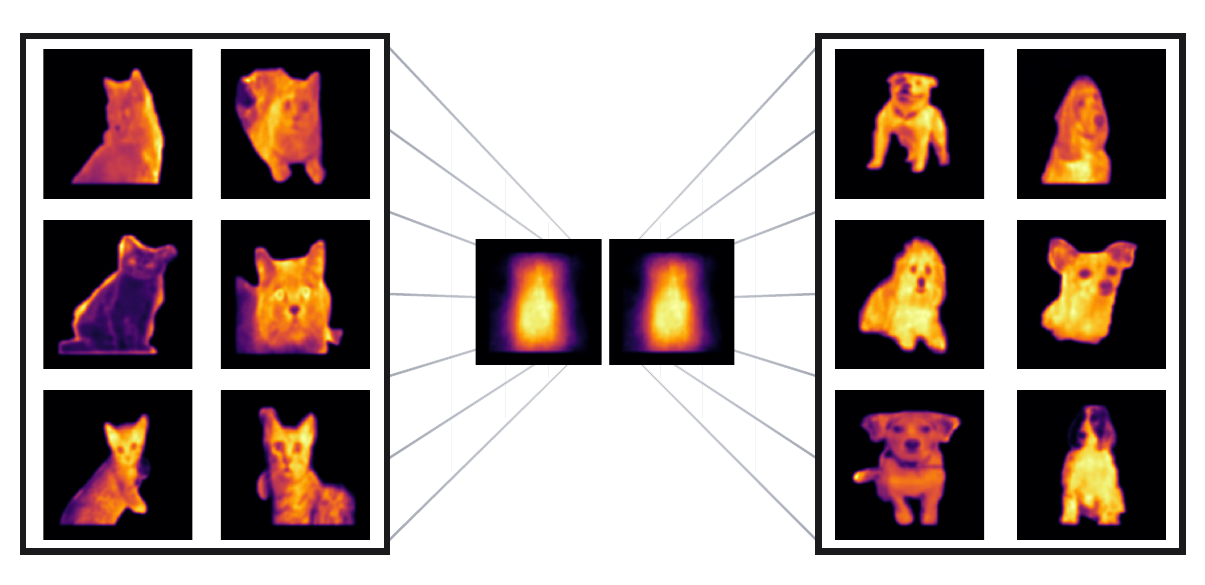}
\caption{\emph{Cats vs dogs} dataset. The pictures are selected so that DI (center) projects the density matrices into almost indistinguishable distributions.}\label{fig:CnD_dataset}
\end{figure}

We start with a binary problem, where we use a dataset of cat and dog images  (Fig.~\ref{fig:CnD_dataset}). The dataset is deliberately constructed such that the ensembles composed of the images of the two classes according to Eq.~\eqref{rho} are almost indistinguishable by imaging, i.e.~$\mathrm{diag}(\hat\rho_1)\simeq\mathrm{diag}(\hat\rho_2)$ in the transverse position basis.

\paragraph{Noiseless detection.}
First we consider the case where photon detection does not suffer from any technical noise. We acquire the DONN output intensity distribution for each instance using a strong light beam and a conventional CMOS camera. Intensity distributions for different instances within the same class are added. To estimate the performance in the photon-counting regime, we then randomly draw simulated photon samples from each resulting distribution. 

\begin{figure*}
    \centering
        \includegraphics[width=0.6\linewidth]{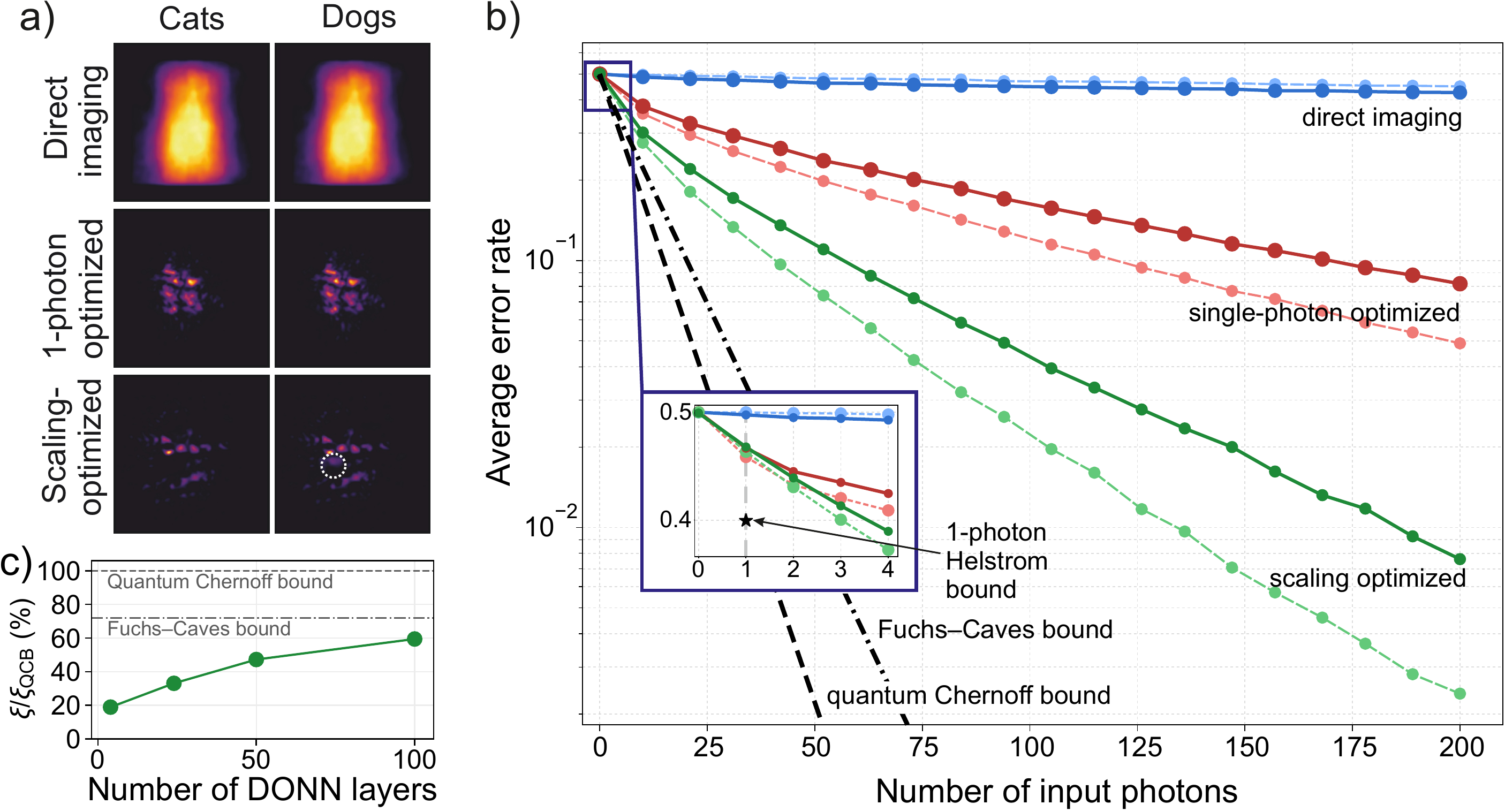}
    \caption{Results for the cats and dogs dataset with noiseless detection.
    (a) DONN output probability distributions in the camera plane. White circle shows a distinguishable feature for the scaling-optimized DONN.
    (b) Average error rate vs. photon number. Solid lines show experimental results, dashed simulations. The inset compares the single-photon errors with the Helstrom bound.
    c) Error-decay exponent versus DONN depth (simulation) for the scaling-optimized training. The Chernoff exponent increases monotonically with the number of phase masks, approaching the Fuchs-Caves bound.%
    }
    \label{fig:CND_classical}
\end{figure*}

Figure \ref{fig:CND_classical}(a) shows the measured output intensity distribution for DI and the DONNs trained using the two methods. Notably, neither the direct-imaging nor the single-photon-optimized distributions are easy to distinguish with the naked eye. In the scaling-optimized case a region of visible difference between the classes emerges (marked with a white circle).

Since the photons are detected independently, the classification error decays exponentially at large photon numbers $N$, $P_\mathrm{err}\propto e^{-\xi(\Phi) N}$, with the classical Chernoff information $\xi(\Phi)$ being bounded from above by the quantum Chernoff exponent $\xi_\mathrm{QCB}$ [Fig.~\ref{fig:CND_classical}(b)], \red{and, in the Bhattacharyya approximation, by the Fuchs-Caves bound $B_\mathrm{FC}=72\% \xi_\mathrm{QCB}$. }

For the optically acquired distributions, DI shows essentially no  decay ($\xi_{\rm DI} \approx 0.5\%\,\xi_\mathrm{QCB}$). %
Both trained measurements decay exponentially, but with considerably different exponents: $\xi(\Phi) = 6\%\,\xi_\mathrm{QCB}$ for the single-photon-optimized DONN and  $16\%$ for the scaling-optimized one. This asymptotic behaviour contrasts with that for $N=1$ (inset), where the single-photon optimized DONN is of advantage. %

Notably, in this case, the scaling-optimized measurement has the lowest photon-detection efficiency: more than 30\% of the incident light is scattered while propagating through the DONN, compared to only $\sim$5\% for the single-photon-optimized case. Because the two states overlap heavily, much of each carries no discriminative information, so the network is free to discard the overlapping parts. Details on how the optical loss is taken into account in training can be found in Methods.

\paragraph{EM-CCD and detection with technical noise.} The advantage demonstrated above assumes noiseless detection. Real-world single-photon detection, however, is always subject to technical noise, whose sources vary between camera types. To study the effects of the noise, we utilize an EM-CCD camera, a standard solution for few-photon imaging in applications ranging from dark-sky astronomy to fluorescence microscopy, operated in photon-counting mode. During the acquisition, the intensity of the laser source
was adjusted so that the effective photon number seen by the camera per frame corresponded to 1 photon at the input of the DONN. The input image from a given class was selected randomly for each acquired frame, thus ensuring the input ensemble is consistent with Eq.~\eqref{rho}. To produce a sample with an average photon number of $N$, we sum $N$ acquired frames, so that the SNR stays constant across photon numbers.

The dominant technical noise is the clock-induced charge, which arises per read-out rather than accumulating with exposure time. At $\sim 0.02$ electrons per pixel per frame, it results in $\sim 10^4$ spurious clicks on the camera sensor per single detected signal photon. %

\begin{figure}
    \centering
        \includegraphics[width=\linewidth]{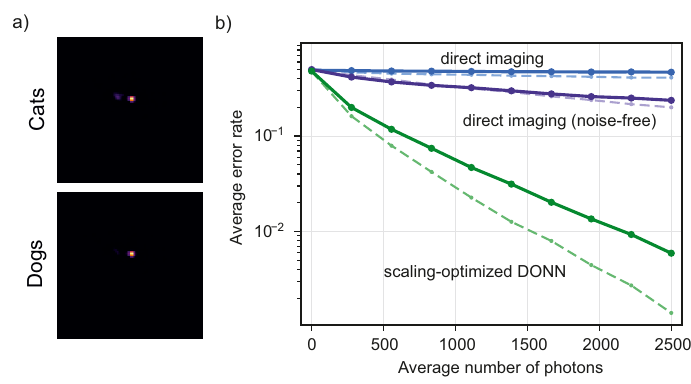}
    \caption{Photon-counting EM-CCD experiment with the effective dark count level of $10^4$ photons per frame. (a) Intensity distributions in the camera plane. Noise-aware training of the scaling-optimized measurement forces the system to focus light into tighter regions, effectively mitigating large portion of noise. (b) Error rate as a function of the average photon number. Solid: experimental, dashed: simulation.}
    \label{fig:CnD_EMCCD}
\end{figure}

Accounting for the noise in scaling-optimized training is discussed in Methods. The resulting DONN compromises between optical state distinguishability and noise mitigation [Fig.~\ref{fig:CnD_EMCCD}(a)]: focusing the light into smaller areas on the camera reduces the number of noise photons in the area where the  output intensity distribution is localized, but focusing too tightly would prevent that distribution from being resolved by the camera pixels. The DONN learns to demagnify the images onto an area of $80\times80$ pixels,
bringing the noise down by two orders of magnitude;
the signal-to-noise ratio nevertheless remains extremely
low making the detection deeply sub-noise. Note that we have not been able to find a reliable method to train the DONN in the single-photon optimized regime in the presence of noise (see Methods).

The DONN error exponent is at a level of $1.4\%$ $\xi_{\rm QCB}$, which is surprisingly high under this deeply sub-noise detection [Fig.~\ref{fig:CnD_EMCCD}(b)]. This rate surpasses the DI error exponent not only for the EMCCD acquisition, but also  $\xi_{\rm DI} \approx 0.5\%\,\xi_\mathrm{QCB}$ obtained in the \emph{noise-free} (classical camera) regime. 
Experimentally acquired DI-based classification with EM-CCD shows no significant difference from random guessing despite demagnification applied in order to reduce the noise effect.

\subsection*{Multi-class discrimination}

Having verified the advantage experimentally on the binary task, we now test how it scales with the number of classes, using the ten-class Fashion-MNIST benchmark \cite{xiao2017fashion}. For more than two classes, the asymptotic error -- and hence the quantum Chernoff bound $\xi_\mathrm{QCB}$ -- is governed by the hardest pair of classes to distinguish: once the easier classes are resolved, the residual error comes from confusing the two closest states \cite{li2016discriminating}. Accordingly, we use $\xi=\min_{i\ne j}\xi_{ij}$, where $\xi_{ij}$ is the Chernoff exponent for classes $i$ and $j$, as our DONN training objective. For training and testing, we randomly selected a subset with 200 images in each category.

\begin{figure}
    \centering
    \includegraphics[width=\linewidth]{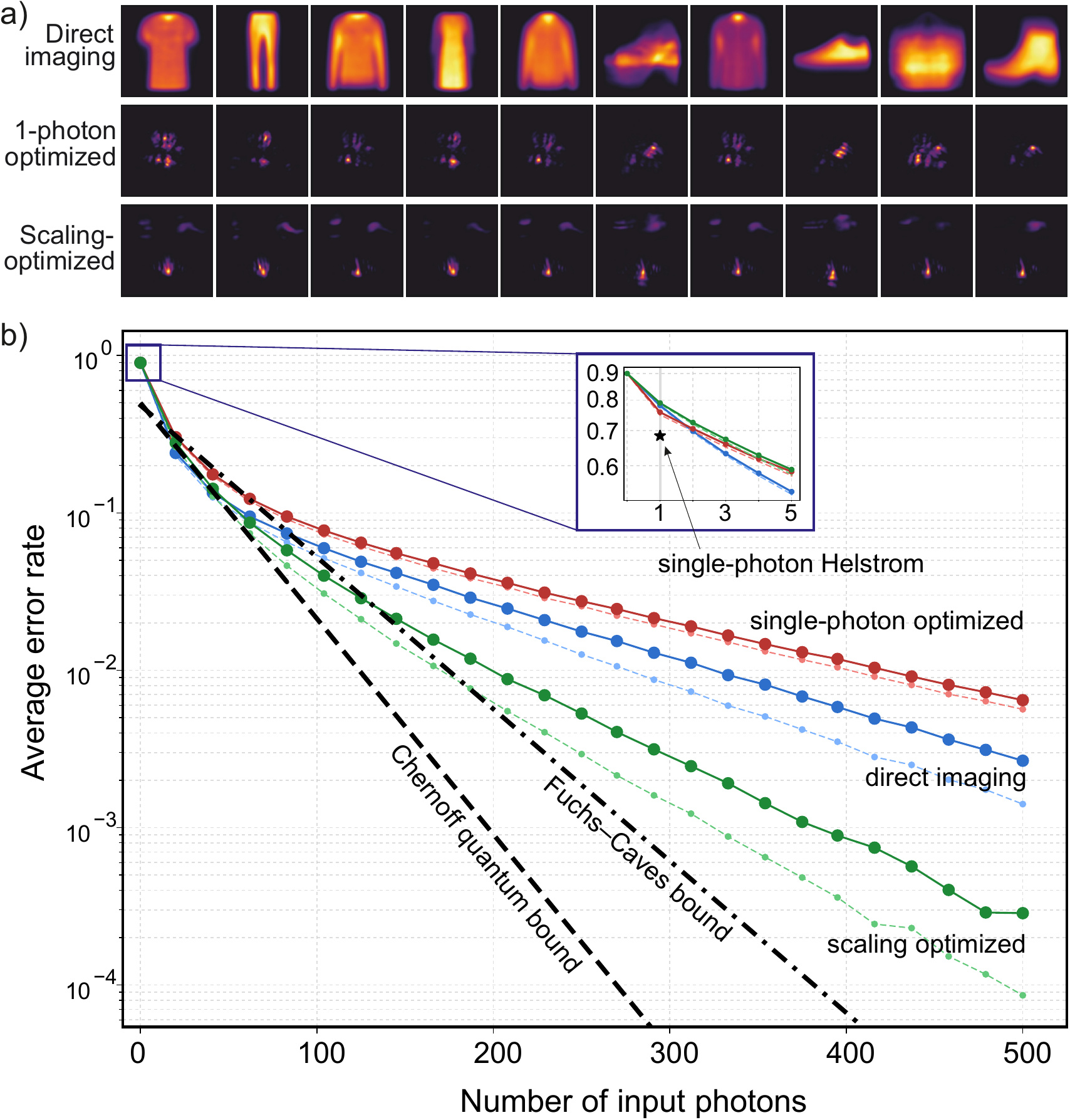}
    \caption{Results for 10-class FMNIST dataset, noiseless detection.
    (a) Intensity distributions in the camera plane.
    (b) Average error rate vs. photon number, evaluated based on the distributions shown in (a).}
    \label{fig:FMNIST_no_noise}
\end{figure}

The experiment was limited to the noise-free regime; the results are shown in Fig.~\ref{fig:FMNIST_no_noise}. Interestingly, the single-photon-optimized measurement, while performing best for $N=1$ [Fig.~\ref{fig:FMNIST_no_noise}(b), inset] obtains the smallest decay exponent of the three ($\xi = 19\%\,\xi_\mathrm{QCB}$), performing more poorly than DI ($28\%$) at high photon numbers. The scaling-optimized measurement, exhibiting $\xi = 42\%\,\xi_\mathrm{QCB}$, performs best at $N\gtrsim60$, where its error falls below DI by $3.3\times$ at $250$ photons and $10\times$ at $500$ photons [Fig.~\ref{fig:FMNIST_no_noise}(b)].

\red{Fig.~\ref{fig:FMNIST_no_noise}(b) also shows the Fuchs–Caves bound of the hardest pair (pullover vs.\ coat), $B_\mathrm{FC}=70\%\,\xi_\mathrm{QCB}$. In the multi-class setting it is an upper bound on the exponent of any individual measurement, but it may not be attainable: a POVM optimal for one pair can make other pairs harder to separate, whereas a single measurement has to distinguish all pairs at once.}

\section*{Discussion}
We demonstrated experimentally that the performance of photon-starved classification is governed by the choice of the optical measurement, and that this measurement can be trained for the task. Central to our results is the choice of the training objective: a measurement optimized for a single detected photon is generally not the one that performs best as photons accumulate, this distinction has been previously observed only for low-dimensional states. Targeting the Chernoff information instead yields exponentially lower error at realistic photon budgets.  To our knowledge, ours is the first experiment to realize a passive optical measurement optimized for photon accumulation: other demonstrations either train the measurement for single-photon accuracy \cite{fan2026quantum} or realize the measurement-side optimization only in simulation \cite{ma2026machine}. The optimization method is capable of accounting for real-world imperfections, such as optical losses and detector noise.

Because the ensembles to be discriminated are known \emph{a priori}, no trainable postprocessing of the detected data is required. This would not be the case if the model were required to generalize to states that were not a part of the training set. In this case, a programmable optical frontend and a digital backend need to be trained jointly, by backpropagating through a stochastic detection model, to maximize classification accuracy \cite{ma2026machine}. Such joint framing blends together the contributions of the optical measurement and the digital neural network. In our study, in contrast, only the measurement front end is variable, letting us unambiguously estimate the advantage arising thanks to the measurement and, in the absence of technical noise, benchmark it against fundamental limits. 

The complexity of visual state discrimination arises from  random location of each single photon detection, resulting in a new random sample every time. If only classical data processing methods were available, generalization would still be required to address this randomness. But optimizing the measurement at the quantum level enables  measurement design as differentiable, sampling-free optimization. %

Our results also show that the most useful measurement is not necessarily the one that minimizes the photon loss within the measurement setup. %
The trained optical system can choose to discard parts of the output images that are identical for different classes since they do not carry any information. 

We also show that with technical noise incorporated into the objective, the measurement finds a balance between separating the states and mitigating the noise.

The relation between trainable optical measurements and the fundamental quantum limits remains to be clarified. Part of the gap to the quantum Chernoff bound is simply a matter of optical expressivity: the four-layer measurement used throughout reaches only about a quarter of the Chernoff exponent in simulation ($19\%$ of $\xi_\mathrm{QCB}$; $16\%$ for the optically acquired distributions), but increasing the DONN depth raises it monotonically toward the bound -- specifically, to $59\%$ at  $100$ layers in simulation [Fig.~\ref{fig:CND_classical}(c)], \red{approaching the Fuchs–Caves limit for individual measurements, 72\% of $\xi_\mathrm{QCB}$. %
The gap between the two bounds can be closed by collective (entangling) measurements over multiple photons.
}

Several extensions follow naturally. First, our work assumes coherent optical fields generated from known input amplitudes; moving to partially coherent or incoherent imaging would bring the framework closer to real-world tasks. Second, a more general computer-vision setting would require classification of previously unseen objects, for which the optical measurement and the digital inference backbone should be trained jointly, since simple maximum-likelihood comparison would no longer be optimal. Third, the same principle can be extended beyond classification: for regression or parameter-estimation tasks, Fisher information provides the natural objective for designing measurements that maximise sensitivity to the relevant physical or semantic parameters. The promise of DONNs in this context --- specifically, for quantum-optimal imaging --- has been shown theoretically \cite{warke2026quantum}.

An important innovation in our work is the introduction of AI methods into the design of quantum measurements. This allows drastic increase of the Hilbert spaces dimensionality accessible to the quantum sensing toolbox: evaluating the ultimate precision limit for estimating a given set of parameters of a given quantum system and constructing instruments that reach this limit. In this way, our approach opens up the quantum dimension in light-starved computer vision and paves the way to reaching the quantum precision in even more complex inference problems.

\begin{acknowledgments}
The project is funded by an EPSRC Standard Grant EP/Y020596/1. We thank Zeyuan (Josh) Luo for designing the cats and dogs dataset. 
\end{acknowledgments}

\section*{Methods}
\subsection*{Experimental setup and digital twin}

To create the objects to be imaged, we use an SLM (SLM$_1$ in Fig.~\ref{fig:setup_simple}) that creates images in the first diffraction order. For each object, the non-negative digital value of each pixel is converted into the amplitude of the optical field. To ensure fair comparison with DI, no information is encoded in the phase of the optical field; the phase is kept constant across each object. 

The generated fields are subsequently demagnified and sent into the 4-layer DONN, comprised by SLM$_2$ and a mirror. Each layer consists of 200 by 200 pixels, each of a size of 9.2 um. The distance between the layers is 8.5 cm, and the distance from the last layer to the camera is 14.3 cm. The wavelength of the laser light is 795 nm.

A practical complication of SLM-based phase modulation is that a fraction of the incident light is reflected without acquiring the programmed phase. To separate this unmodulated component from the signal, each DONN layer displays, on top of its trained phase mask, a horizontal blazed grating with a period of 4 pixels so the modulated light is deflected into the first diffraction order, where the next layer is placed. The unmodulated light propagates undeflected and leaves the optical path. The price of such a short-period grating is its limited diffraction efficiency, with losses accumulating to roughly 20~dB over the 4 bounces. We exclude these losses from the photon-budget accounting, since they are specific to the SLM implementation and would be absent for physically manufactured phase plates \cite{labroille2014efficient, chen2021diffractive}. We chose to use an SLM, rather than fixed holograms, for its reconfigurability, which allows all measurements to be compared in the same setup.

To train the diffractive-optical measurement, we use a digital twin simulated via TorchOptics  package \cite{filipovich2024torchoptics}. Each element of the dataset is propagated through the DONN, after which the transverse intensity distribution of the field is calculated. Based on the distributions produced by different samples and depending on the final objective, the loss function value is obtained, which is then back-propagated through the system to update the phase values for each pixel in each phase plane.

\subsubsection*{State normalization and the loss pixel}
A na\"ive choice is to normalize the output DONN distributions entering Eq.~\eqref{eq:bhattacharyya} to unity irrespective of the loss. However, the resulting training objective would let the diffractive setup scatter away as much light as it likes to make the surviving distributions look different, whereas our photon budget is fixed at the \textit{input} of the measurement. We therefore normalize with respect to the total input intensity, so that $P_1(x)$ and $P_2(x)$ generally sum to less than one, with the deficit corresponding to light scattered out of the detected region.

To keep the distributions normalized we associate this scattered light with a single \textit{loss pixel} (lp): $P_i(lp) = 1 - \sum_{x}P_i(x)$. The Bhattacharyya objective then reads
\begin{equation}
B(\Phi) = - \log  \sum_{x'\in\{x\}\cup \{{\rm lp}\}} \sqrt{P_1(x')\, P_2(x')}.
\label{eq:bhattacharyya_w_loss}
\end{equation}

This construction discourages scattering implicitly rather than through an arbitrary penalty term. No matter how light is scattered, it all lands in the same loss pixel, so lost light becomes indistinguishable between the two classes and reduces the score. On the other hand, if instead of discarding light the system can redistribute it over camera pixels even slightly differently for the two distributions, it gains score  \eqref{eq:bhattacharyya_w_loss} by doing so. The training reinforces such redistribution. 

The construction also captures a useful distinction. Suppose a camera pixel $x_1$ carries equal probability for both classes, $P_1(x_1) = P_2(x_1)$. A photon detected there conveys no class information, so detecting it is equivalent to losing it; moving that light to the loss pixel leaves the objective \eqref{eq:bhattacharyya_w_loss} unchanged. The training thus penalizes only the loss of \emph{distinguishable} light, not the loss of uninformative light. This is consistent with our observation that scaling-optimized measurements achieve the best accuracy despite having the highest scattering loss.

\subsubsection*{Training with noise}
Real single-photon cameras produce a significant number of spurious counts. We model this as Poisson noise with a mean count per pixel that is independent of the signal. We assume the noise to be uniform across the camera and write its contribution as a constant baseline $n$ added to every pixel. %
Every recorded click is then a sample from the combined signal-plus-background distribution. The noise background can therefore be treated as part of the state, and the task becomes the discrimination between the effective distributions $P_i(x \mid \Phi) + n$. Adding this baseline to Eq.~\eqref{eq:bhattacharyya} gives
\begin{equation}
B(\Phi)= - \log \left[ \sum_{x} \sqrt{\big(P_1(x)+n\big)\big(P_2(x)+n\big)} \right ].
\label{eq:bhattacharyya_noise}
\end{equation}
An important feature of this objective is that the score improves when the distributions are concentrated on fewer pixels. Localizing the signal effectively excludes the empty, noise-only pixels from consideration, raising the signal-to-noise ratio.

The single-photon objective \eqref{eq:single_photon_acc} has no natural counterpart to this construction: a uniform baseline added to every class leaves the per-pixel maximum, and hence the decision map, unchanged, contributing only a near-constant offset to the objective. In particular, it creates no incentive to concentrate the signal, since the accuracy of a single detected count does not depend on how the signal is spread over the camera. The benefit of noise mitigation arises only as evidence accumulates over multiple detections. %

Combining the loss pixel with noise requires care in how much noise to assign to that pixel. If it carries little or no noise, the loss pixel becomes a haven: the training can improve the apparent score by scattering light differently for the two classes into a pixel we never actually read out, which does not serve the real objective. To prevent this, scattering into the loss pixel must be made at least as costly as spreading light uniformly over the whole camera, which we ensure by assigning it the summed noise of all camera pixels, $n_{lp} = \sum_x n$. The objective with both loss and noise then reads

\begin{equation}
\begin{split}
B(\Phi)= &- \log \Bigg[ \sum_{x} \sqrt{\big(P_1(x)+n\big)\big(P_2(x)+n\big)}  \\ &+\sqrt{\big(P_1(lp)+n_{lp}\big)\big(P_2(lp)+n_{lp}\big)} \Bigg]
\label{eq:bhattacharyya_noise_lp}
\end{split}
\end{equation}

As before, this leaves uninformative light untouched --- the score is unchanged whether a pixel with $P_1(x_1)=P_2(x_1)$ is a camera pixel or the loss pixel --- while making the loss of distinguishable light costly.

\subsubsection*{Beyond binary classification}
For more than two classes, the asymptotic performance is limited by the \textit{hardest} pair to distinguish -- the pair with the smallest pairwise distance. Because the distance is computed over the output distributions, which depend on the phase masks $\Phi$, the hardest pair changes during training. Our objective at each step is therefore to maximize the minimum pairwise Bhattacharyya distance over all class pairs, $\min_{i<j} B(P_i, P_j \mid \Phi)$. To keep this differentiable we replace the hard minimum with a Softmin at temperature $\tau$,
\begin{equation}
\mathcal{L}_{\mathrm{multiclass}}
=
\tau
\log
\left[
\sum_{i<j}
\exp\!\left(
-\frac{B(P_i, P_j \mid \Phi)}{\tau}
\right)
\right],
\label{eq:multiclass_loss}
\end{equation}
where each $B(P_i, P_j \mid \Phi)$ is the loss- and noise-aware Bhattacharyya distance \eqref{eq:bhattacharyya_noise_lp}. %

\bibliography{sn-bibliography}

\end{document}